\documentclass{article}
\usepackage[T1]{fontenc}
\usepackage[utf8]{inputenc}
\usepackage{ismir} 
\usepackage{amsmath,cite,url}
\usepackage{graphicx}
\usepackage{color}
\usepackage[bookmarks=false]{hyperref}

\newcommand{\blue}[1]{#1}
\title{Versatile Symbolic Music-for-Music Modeling via Function Alignment}

\multauthor
{Junyan Jiang$^{1, 2}$ \hspace{1cm} Daniel Chin$^{1, 2}$ \hspace{1cm} Liwei Lin$^{1, 2}$ \hspace{1cm} Xuanjie Liu$^{2}$ \hspace{1cm} Gus Xia$^{1, 2}$}{
$^1$ NYU Shanghai \hspace{1cm}$^2$ Music X Lab, MBZUAI\\
{\tt\small \{jj2731, daniel.chin, ll4270, gxia\}@nyu.edu, xuanjie.liu@mbzuai.ac.ae}
}

\def\authorname{J. Jiang, D. Chin, L. Lin, X. Liu and G. Xia}

\begin{document}

\maketitle

\begin{abstract}
Many music AI models learn a map between music content and human-defined labels. However, many annotations, such as chords, can be naturally expressed within the music modality itself, e.g., as sequences of symbolic notes. This observation enables both understanding tasks (e.g., chord recognition) and conditional generation tasks (e.g., chord-conditioned melody generation) to be unified under a music-for-music sequence modeling paradigm. In this work, we propose parameter-efficient solutions for a variety of symbolic music-for-music tasks. The high-level idea is that (1) we utilize a pretrained Language Model (LM) for both the reference and the target sequence and (2) we link these two LMs via a lightweight adapter. Experiments show that our method achieves superior performance among different tasks such as chord recognition, melody generation, and drum track generation. All demos, code and model weights are publicly available\footnote{\blue{\href{https://github.com/music-x-lab/midi-function-alignment}{https://github.com/music-x-lab/midi-function-alignment}}}. 

\end{abstract}

\section{Introduction}\label{sec:introduction}



Many foundational tasks in music AI, such as music information retrieval (MIR) and conditional music generation, have traditionally been formulated as mappings between music and labels: either from music to task-specific annotations (e.g., chord recognition), or from descriptive conditions to music (e.g., chord-conditioned melody generation). While these tasks have long been treated separately, a key observation is that in many cases, the “labels” themselves can also be represented in the \textit{same music modality}—for example, as note sequences. This suggests a unifying perspective: a wide range of MIR and generation tasks can be reformulated as sequence-to-sequence problems within the music domain. We refer to this formulation as \textit{music-for-music}\textbf{ }modeling.

\begin{figure}
  \centering
  \includegraphics[width=0.9\linewidth, page=3, clip, trim=1.4cm 12cm 11.9cm 0.8cm]{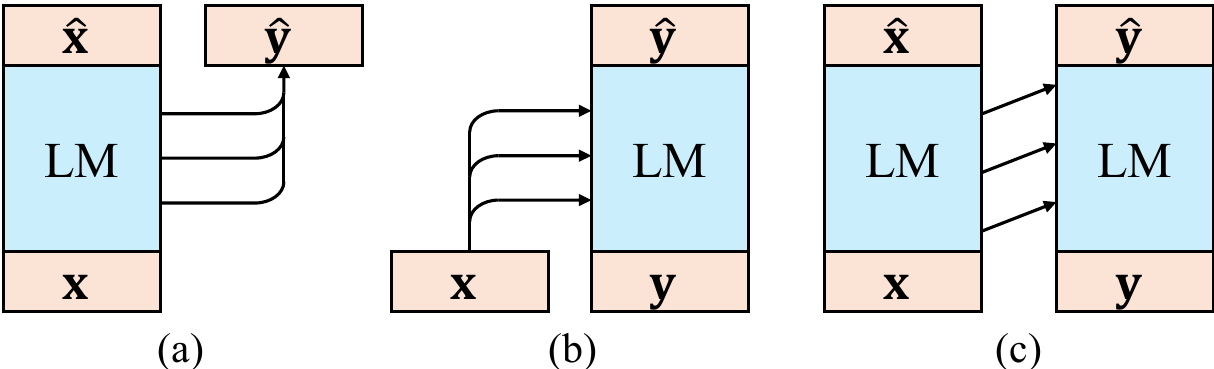}
  \caption{Three types of sequence-to-sequence models by knowledge transfer from pretrained LMs. $\mathbf{x}$ and $\mathbf{y}$ are input sequences and $\hat{\mathbf{x}}$ and $\hat{\mathbf{y}}$ are predictions, possibly right-shifted due to the autoregressive targets. (a) Probing; (b) Prefix tuning; (c) Function alignment ($\mathbf{x}\rightarrow \mathbf{y}$).}
  \label{fig:function_alignment}
\end{figure}

To achieve versatile music-for-music modeling in a sample-efficient way, we apply \textit{knowledge transfer} to pretrained foundational Language Models (LMs) using a light-parameterized adaptor. As illustrated in Fig.~\ref{fig:function_alignment}(a)-(b), many existing methods such as probing~\cite{donahue2022melody,mert,mertech} and prefix tuning~\cite{li2021prefix,zhang2023llama,cocomulla} transfer knowledge of foundation models to downstream tasks by adapting them to new input or output,  but the knowledge resides in only one language---either the LM of source  $\mathbf{x}$ or the LM of target $\mathbf{y}$. In contrast, our method distills knowledge from \textit{both} LMs via aligning them in a layer-wise manner, as shown in  Fig.~\ref{fig:function_alignment}(c). 

At the methodology level, our approach is inspired by \textit{function alignment}~\cite{function_alignment}, a recently proposed theory of mind that attributes the emergence of intelligence to the dynamic synergy among interacting agents\blue{, i.e., Language Models (LMs)}. In our work, we contribute two concrete implementations of this idea—\blue{by} creating synergy \blue{between two LMs} through Parameter-Efficient Fine-Tuning (PEFT).

The first approach introduces a trainable cross-attention layer between two separately pretrained LMs. The second, more concise solution, uses a lightweight self-attentive adapter applied to concatenated input-output sequences within a \textit{single shared LM}—a strategy applicable when both input and output share the same vocabulary. We show the effectiveness of both implementations using experiments on both generative and analysis tasks, including: (1) chord-conditioned melody generation, (2) melody-conditioned chord generation, (3) drum-conditioned song generation, (4) song-conditioned drum generation and (5) few-shot symbolic music analysis.

The main contribution of this paper is as follows:
\begin{enumerate}\setlength\itemsep{0em}
    \item We achieve versatile \textit{music-for-music} modeling, unifying a broad range of music understanding and controllable generation tasks under a shared framework.
    
    \item At the methodological level, we bring the novel concept of function alignment---a recently proposed theory of mind that emphasizes synergy among agents---into the domain of music AI, offering a fresh perspective on sequence-to-sequence tasks.

    \item While the original position paper on function alignment remains at a conceptual level, our work takes a significant step forward by introducing two concrete, \textit{parameter-efficient} implementations in the context of modern language models: one via cross-attentive adapters across two LMs, and another via a self-attentive adapter within a shared LM. We demonstrate the effectiveness of both approaches through theoretical analysis and empirical validation.

\end{enumerate}


\section{Related Works}

\subsection{Music Foundation Models}

Since the invention of the Transformer architecture~\cite{vaswani2017attention}, transformer-based language models have become the mainstream of music foundation models on multiple modalities, including audio~\cite{dhariwal2020jukebox,musicgen,musiclm,zhang2025inspiremusic}, symbolic~\cite{musictransformer,musecoco,huang2020pop,le2024meteor,chensympac,zhang2025generating,shih2022theme,qu2024mupt,shu2024musebarcontrol} and text-based music representation~\cite{chatmusician}. In addition to autoregressive models, masked language models~\cite{mert,garcia2023vampnet} and diffusion models~\cite{chen2024musicldm,wu2024music,lam2023efficient,schneider2024mousai,hou2025editing,wang2024whole} and flow-based models~\cite{tal2024joint} can also be used as foundation models, but we focus on autoregressive models in the literature review.


For symbolic music, \blue{the} music transformer~\cite{musictransformer} is an early work to adopt the transformer architecture to music. Some follow-up works try to design a better representation of the music content. For example, pop music transformer imposes a metrical structure in the data representation~\cite{huang2020pop}. MuPT trains transformers on their proposed synchronized multi-track ABC notation~\cite{qu2024mupt}. Other works aim to introduce controllability to the generative model. \blue{MuseCoco} generates the music score from text~\cite{musecoco}. 
METEOR performs melody-aware orchestral music generation with texture control~\cite{le2024meteor}. 
SymPAC trains symbolic generation models from transcribed audio data with chord, section, and instrument controls~\cite{chensympac}. 
Zhang \textit{et al.} improve generation discriminators to better follow rhythm and melody conditions ~\cite{zhang2025generating}. 
The Theme Transformer~\cite{shih2022theme} uses a short theme condition for generation. MuseBarControl generates music with fine-grained control to the bar level~\cite{shu2024musebarcontrol}. 

\subsection{Parameter-Efficient Fine-Tuning}
Parameter-Efficient Fine-Tuning (PEFT) methods add lightly parameterized adapters to large pretrained models. Compared to full-parameter fine-tuning, PEFT requires significantly less computation and training data. Existing methods include appending task-specific prefixes to input sequences~\cite{li2021prefix,liu2022p}, injecting low-rank adaptation (LoRA) to linear layers~\cite{hu2022lora}, and adding learnable hidden states to the self-attention blocks~\cite{zhang2023llama,gao2023llama}. 

PEFT has been applied to music foundation models to support new tasks. 
Coco-Mulla~\cite{cocomulla} and MusiConGen\cite{lanmusicongen} both adapt MusicGen to follow content controls such as chord and rhythm. 
Additionally, AirGen enables MusicGen to infill segments based on content controls~\cite{airgen}. 
Instruct-MusicGen extends MusicGen for music editing by text instructions~\cite{zhang2024instruct}. 
Audio Prompt Adapter extends AudioLDM2 for music editing following controls such as genre, timbre, and melody~\cite{tsai2024audio}. 
Ou \textit{et al.} \blue{tunes} a symbolic language model for \blue{tasks like} band arrangement, piano reduction, drum arrangement and voice separation~\cite{longsan}. 

\section{Methodology}

\subsection{Base Model}

For this study, we choose the base model (the pretrained symbolic LM) with two main considerations. First, we do not wish to introduce \textit{any} control in the pretraining stage, since we want to demonstrate the controllability using PEFT. We refain from using any annotation or metadata (i.e., chord, bar or text annotations) to pretrain the base model. Second, we want to adopt a data representation that can help the model align multiple sequences in time easily. Instead of using a MIDI event-like representation~\cite{musictransformer,huang2020pop,hsiao2021compound} where two time-aligned sequences might have a significant length difference, we use a fixed time step (a 16$^\textrm{th}$ note unit) for the input sequence.

\begin{figure}[t]
  \centering
  \includegraphics[width=0.9\linewidth, page=6, clip, trim=0.2cm 6.4cm 17.6cm 0.9cm]{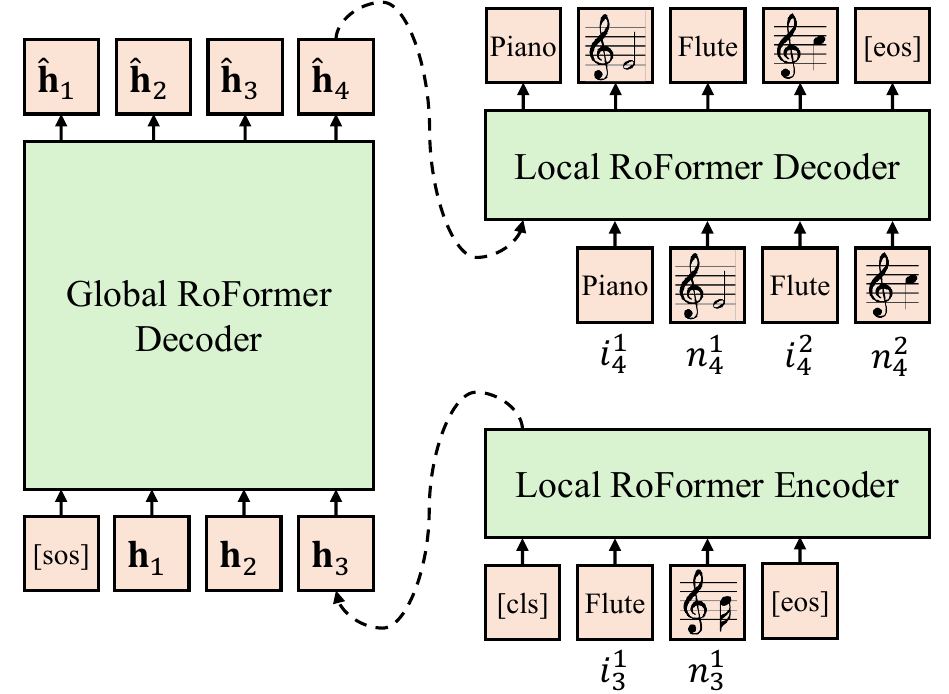}
  \caption{The architecture of the foundation model. The left side shows the global decoder. The right side shows the encoding of a single time step $\textbf{x}_3=\{i^1_3,n^1_3, \textrm{[eos]}\}$ and the decoding of the next step $\textbf{x}_4=\{i^1_4,n^1_4,i^2_4,n^2_4,\textrm{[eos]}\}$.}
  \label{fig:base_model}
\end{figure}

Since multiple notes can occur at the same time step, we use a hierarchical scheme to compress (decompress) the note lists on the same time step with a local encoder (decoder), as shown in Fig.~\ref{fig:base_model}.

\subsubsection{Data Representation}

Formally, we represent a score sequence $\mathbf{x}=\{\mathbf{x}_1, ...,\mathbf{x}_T\}$ with a fixed time step of a 16$^\textrm{th}$ note. Since each time step may contain multiple note onsets, each $\mathbf{x}_t$ represents a list of $N_t$ notes whose quantized onset time is the $t$-th 16$^\textrm{th}$ note (i.e., $\textbf{x}_t$ is a simu-note~\cite{wang2020pianotree} at time step $t$). We define
\begin{equation} \label{eqn:sublist}
    \mathbf{x}_t=\{i_t^1, n_t^1,i_t^2, n_t^2,...,i_t^{N_t},n_t^{N_t}, \textrm{[eos]}\}
\end{equation}
where \blue{$i_t^k\in\{0,...,128\}$} is the instrument ID for the $k$-th note. We use the MIDI program number $0...127$ for pitched instruments and $i_t^k=128$ for drums. $n_t^k=24p_t^k+d_t^k$ is a flattened representation of the $k$-th note's pitch \blue{$p_t^k\in\{0,...,127\}$ and duration $d_t^k\in\{0,...,23\}$}. $p_t^k$ denotes the MIDI pitch from 0 to 127. $d_t^k\in\{0,...,23\}$ is the note duration quantized into 24 possible bins, $d_t^k=j$ corresponds to a duration of $b_j$ sixteenth notes where $\mathbf{b}=[1, 2, 3, 4, 6, 8, 12, 16, 24, ..., 4096]$. $\textrm{[eos]}$ is a special token marking the end of the list. All notes in $\mathbf{x}_t$ are sorted primarily by $i_t^k$ and secondarily by $n_t^k$.

\subsubsection{Model Design}

We use a RoFormer~\cite{su2024roformer}, a popular transformer architecture as the backbone model. The model architecture is shown in Fig.~\ref{fig:base_model}. Since our input sequence contains nested lists, we first encode each $\mathbf{x}_t$ with a local RoFormer encoder:
\begin{equation}\label{eqn:local_enc}[\mathbf{h}_t,\_]=\textrm{LocalEncoder}(\textrm{[cls]},\mathbf{x}_t)
\end{equation}
for all $t=1...T$. Specifically, we prepend a [cls] token at the beginning of $\mathbf{x}_t$ and pass the sequence to the encoder. $\mathbf{h}_t$ is acquired from the output representation of the [cls] token. We then use a global RoFormer decoder to autoregressively model the symbolic score:
\begin{equation}\label{eqn:global_dec}
\hat{\mathbf{h}}_t=\textrm{GlobalDecoder}(\mathbf{e}_{\textrm{sos}}, \mathbf{h}_{1...t-1})
\end{equation}
where $\mathbf{e}_{\textrm{sos}}$ is a learnable start-of-sentence (sos) embedding. Finally, a local RoFormer decoder generates each note by
\begin{equation}
\hat{\mathbf{x}}_{t, j}=\textrm{LocalDecoder}(\label{eqn:local_dec}\hat{\mathbf{h}}_t,\mathbf{x}_{t, 1...j-1})
\end{equation}
for all $t=1...T$. Here, $\mathbf{x}_{t,j}$ denotes the $j$-th token of list $\mathbf{x}_{t}$ (see Eqn.~\ref{eqn:sublist}). The local decoder terminates when an end-of-sentence (eos) token is generated.

We will use $\mathbf{\hat{x}}_t=\textrm{LM}(\mathbf{x}_{0...t-1})$ (or simply $\textrm{LM}(\mathbf{x})$) as a shorthand for the autoregressive model of sequence $\mathbf{x}$ through Eqs.~\ref{eqn:local_enc}-\ref{eqn:local_dec}. Here, $\mathbf{x}_0$ denotes the global start-of-sentence embedding $\mathbf{e}_\textrm{sos}$.


\begin{figure}[t]
  \centering
  \includegraphics[width=0.9\linewidth, page=4, clip, trim=0.2cm 6.0cm 12.6cm 0.5cm]{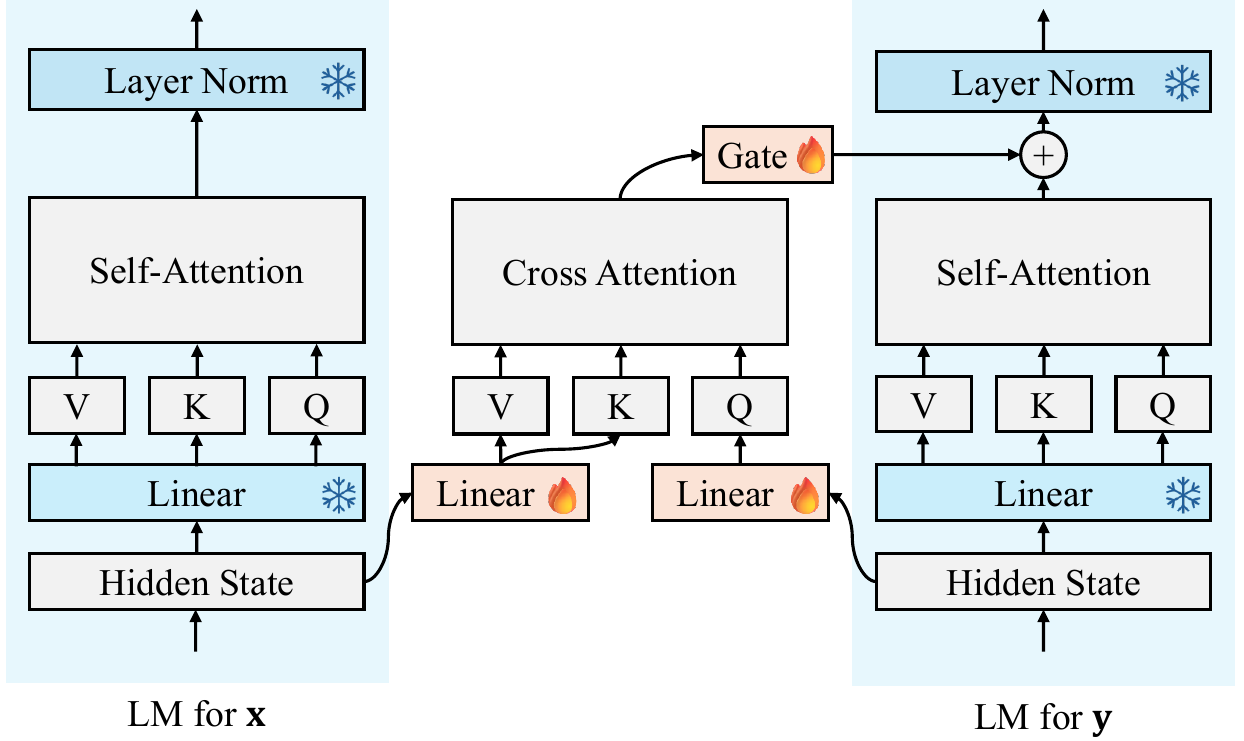}
  \caption{The architecture of a cross-attentive function alignment adapter. The fire icon denotes trainable parameters, and the snowflake icon denotes frozen parameters.}
  \label{fig:cross_alignment}
\end{figure}

\begin{figure}[t]
  \centering
  \includegraphics[width=0.7\linewidth, page=5, clip, trim=0.2cm 4.0cm 18.5cm 0.3cm]{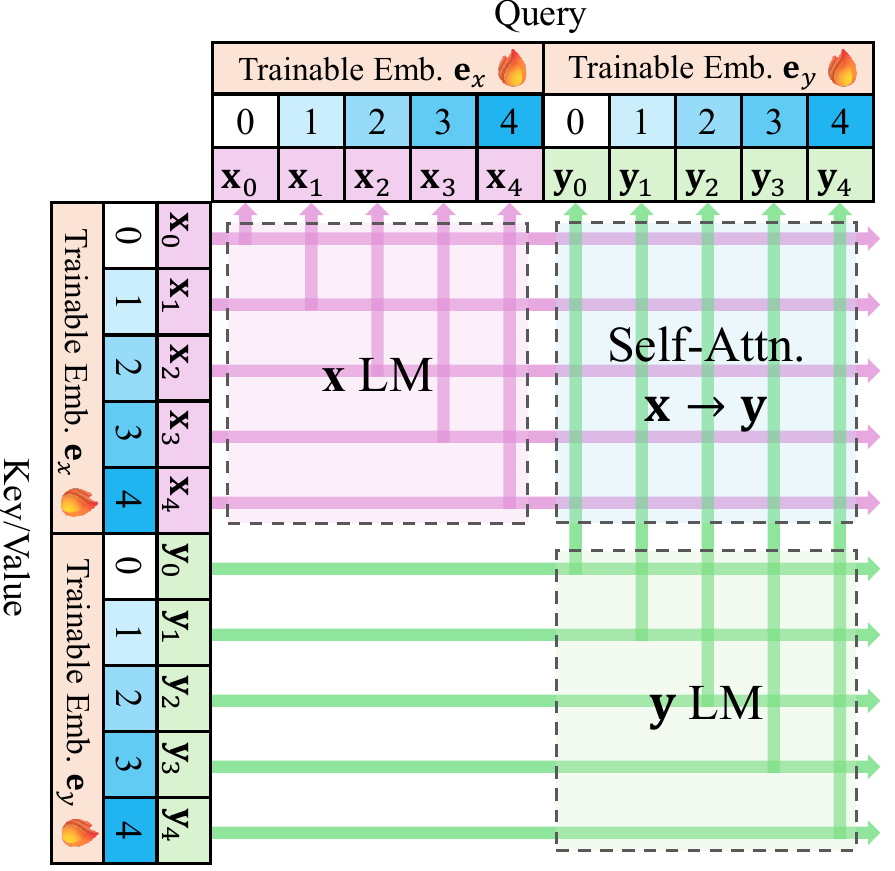}
  \caption{The architecture of a self-attentive function alignment adapter. Crossed vertical and horizontal arrows indicate the flow of information between the corresponding query and key/value pairs, while all other connections are masked by the autoregressive self-attention mechanism. The indices 0 through 4 represent the proposed positional embeddings for the concatenated sequence.}
  \label{fig:self_alignment}
\end{figure}

\subsection{Parameter-Efficient Fine-Tuning}

Our fine-tuning strategy leverages pretrained LMs for $\mathbf{x}$ and $\mathbf{y}$, connected via a parameter-efficient module. We present two variants: cross-attentive adapters for separate LMs, and self-attentive adapters for a shared LM. We apply both adapters to the backbone of the foundation model (the global decoder in Eqn.~\ref{eqn:global_dec}) only. 

\subsubsection{Cross-attentive Function Alignment}

Our first approach is to use a cross-attention layer between the hidden layers of two LMs. A similar architecture has been adopted in language processing~\cite{bansal2024llm} and speech processing~\cite{zayats2024zipper}. We refer to the design of \cite{bansal2024llm} and show an adapted version in Fig.~\ref{fig:cross_alignment}. For the $l$-th attention layer of $\textrm{LM}(\mathbf{y})$, the original self-attention is defined as:
\begin{equation}
    \mathbf{h}_{\mathrm{p}}^{l}=\textrm{SelfAttn}(\mathbf{W}_q^{l}\mathbf{z}_y^{l},\mathbf{W}_k^{l}\mathbf{z}_y^{l},\mathbf{W}_v^{l}\mathbf{z}_y^{l})
\end{equation}
where $\mathbf{z}_y^{l}$ denote the $l$-th layer hidden states for $\textrm{LM}(\mathbf{y})$ and $\mathbf{W}^{l}$ denotes pretrained weights. The adapted version can be written as:
\begin{equation} \label{eqn:yinyang_has_g}
    \mathbf{h}_{\mathrm{a}}^{l}=\mathbf{h}_{\mathrm{p}}^{l}+g^l\cdot \textrm{CrossAttn}(\mathbf{U}_q^{l}\mathbf{z}_y^{l},\mathbf{U}_k^{l}\mathbf{z}_x^{l},\mathbf{U}_v^{l}\mathbf{z}_x^{l})
\end{equation}
where \blue{$g^l$} is a zero-initialized trainable gate scaler. $\mathbf{U}^l$ are trainable parameters. Intuitively, this allows the query from $\textrm{LM}(\mathbf{y})$ to attend both to itself (self-attention) and to the condition from $\textrm{LM}(\mathbf{x})$ (cross-attention).

Besides the trainable cross-attention module, we also \blue{apply LoRA}~\cite{hu2022lora} to all $\mathbf{W}_q^{l}$ and $\mathbf{W}_v^{l}$ of both pretrained models $\textrm{LM}(\mathbf{x})$ and $\textrm{LM}(\mathbf{y})$, allowing the model to learn distinctive features of sequences $\mathbf{x}$ and $\mathbf{y}$.

\subsubsection{Self-attentive Function Alignment}

When $\mathbf{x}$ and $\mathbf{y}$ share the same pretrained LM, alignment becomes a special case: it can be achieved by concatenating their sequences and feeding them into a single model. The LM will first model $\mathbf{x}$ and predict $\mathbf{y}$ given $\mathbf{x}$ as a prefix.

This implies that some prior PEFT methods~\cite{airgen,longsan}, which structure the condition and generated sequence within a single language model, can be viewed as broader forms of function alignment. We show that a simpler configuration is also effective and explain why it realizes function alignment.

When we directly concatenate two sequences $[\mathbf{x},\mathbf{y}]$ and feed them to the decoder self-attention layer, we can decompose it into the self-attention of $\mathbf{x}$ and $\mathbf{y}$, and an extra component influencing $\mathbf{y}$ from $\mathbf{x}$, as shown in Fig.~\ref{fig:self_alignment}. Specifically, we have
\begin{equation} \small
\begin{aligned} \label{eqn:joint_xy}
    [\mathbf{h}_{x\mathrm{a}}^{l},\mathbf{h}_{y\mathrm{a}}^{l}]&=\textrm{SelfAttn}(\mathbf{W}_q^{l}[\mathbf{z}_x^{l},\mathbf{z}_y^{l}],\mathbf{W}_k^{l}[\mathbf{z}_x^{l},\mathbf{z}_y^{l}],\mathbf{W}_v^{l}[\mathbf{z}_x^{l},\mathbf{z}_y^{l}]) \\
    &=\textrm{SelfAttn}([\mathbf{Q}_x^{l},\mathbf{Q}_y^{l}],[\mathbf{K}_x^{l},\mathbf{K}_y^{l}],[\mathbf{V}_x^{l},\mathbf{V}_y^{l}])
\end{aligned}
\end{equation}
for every layer $l$. In a single-head setting, we have $\textrm{SelfAttn}(\mathbf{Q},\mathbf{K},\mathbf{V}):=\textrm{softmax}(\mathbf{Q}\mathbf{K}^\top/\sqrt{d}+\mathbf{M})\mathbf{V}$, where $d$ is the dimension of the key vectors and $\mathbf{M}$ is the autoregressive mask. We can rewrite Eqn.~\ref{eqn:joint_xy} by
\begin{equation}\small
\mathbf{h}_{x\mathrm{a}}^{l}=\textrm{SelfAttn}(\mathbf{Q}_x^{l},\mathbf{K}_x^{l},\mathbf{V}_x^{l})
\end{equation}
\begin{equation}\small\begin{aligned}\label{eqn:awful_eqn}
\mathbf{h}_{y\mathrm{a}}^{l}&=\frac{\mathbf{a}}{\mathbf{a}+\mathbf{b}}\textrm{SelfAttn}(\mathbf{Q}_y^{l},\mathbf{K}_y^{l},\mathbf{V}_y^{l})\\
&+\frac{\mathbf{b}}{\mathbf{a}+\mathbf{b}}\textrm{CrossAttn}(\mathbf{Q}_y^{l},\mathbf{K}_x^{l},\mathbf{V}_x^{l})
\end{aligned}\end{equation}
where $\mathbf{a}=\sum_j \exp{\left[(\mathbf{Q}^{l}_{y}\mathbf{K}^{l}_{y})_j/\sqrt{d}+\mathbf{M}\right]}$ and $\mathbf{b}=\sum_j \exp{\left[(\mathbf{Q}^{l}_{y}\mathbf{K}^{l}_{x})_j/\sqrt{d}\right]}$. Note that Eqn.~\ref{eqn:awful_eqn} closely mirrors Eqn.~\ref{eqn:yinyang_has_g} in form. Although the gating vectors $\mathbf{a}$ and $\mathbf{b}$ are not explicitly parameterized, we hypothesize that this design remains effective.

After concatenating $\mathbf{x}$ and $\mathbf{y}$, we reset $\mathbf{y}$'s positional embeddings to start from 0 to better preserve the pretrained behavior of $\textrm{LM}(\mathbf{y})$. To avoid token indistinguishability due to overlapping positions, we also add zero-initialized, learnable sentence embeddings $\mathbf{e}_x$ and $\mathbf{e}_y$ to their respective positional encodings, as shown in Fig.~\ref{fig:self_alignment}.

Similar to cross-attentive adapters, a trainable LoRA module is also appended to the pretrained LM.

\section{Experiments}\label{sec:experiments}

In the experiments, we first describe the hyperparameters and the pretraining scheme of our foundation model (Sec.~\ref{sec:model_pretraining}). We evaluate our adapters on both generative and analysis tasks. We describe the tasks in Sec.~\ref{sec:tasks} and models in Sec.~\ref{sec:baselines}. We then show the setting for subjective evaluation (Sec.~\ref{sec:subjective}) and objective evaluation (Sec.~\ref{sec:objective}), and analyze the results in Sec.~\ref{sec:results}.

\subsection{Model Pretraining} \label{sec:model_pretraining}

We use a RoFormer with a 12-layer global decoder (hidden size 768, intermediate size 3072, 12 heads). The local encoder and decoder are smaller 3-layer RoFormers (hidden size 768, intermediate size 768, 8 heads).

We pretrain our foundation model on the Los Angeles MIDI dataset~\cite{lev2024losangelesmididataset}, which contains approximately 405,000 MIDI files. As a score-based model, it relies on accurate beat annotations (inferred from tempo change events) for correct quantization. However, many files in the pretraining dataset contain incorrect tempo information.

To address this, we apply a rule-based filter. Normally, note onsets are not uniformly distributed across odd and even time steps. We compute the ratio of notes quantized to odd vs. even time steps. If the ratio falls within $0.5 \pm 0.15$ for every track, we assume it is poorly quantized and discard the song. This yields a cleaned subset of 357,279 files. During pretraining, we also apply a random pitch shift within $[-5, 6]$ semitones for data augmentation.

We set the global sequence length to $T=384$ and cap the maximal polyphony by $N_t \leq 16$, clipping excess notes per time step. A batch size of 48 is used for pretraining. We train the model for 2,000,000 iterations using AdamW~\cite{loshchilov2017decoupled} with $\beta{=}(0.9, 0.999)$ and weight decay 0.01. We use a OneCycleLR~\cite{smith2019super} scheduler with a maximum LR $10^{-4}$ and 10,000 warm-up steps. Pretraining takes around 12 days on 4$\times$A100 (40GB) GPUs.

\subsection{Downstream Tasks} \label{sec:tasks}

We evaluate the adaptor on different music generation and understanding task. Specifically, we have 3 sets of tasks:

\begin{itemize}
\item \textbf{Melody to chord} and \textbf{chord to melody}: we fine-tune the model on the Nottingham dataset~\cite{nottingham} with a total of 1,020 songs. The model is asked to generate chords from a given melody or to generate a melody given a chord progression.
\item \textbf{Drum to others} and \textbf{others to drum}: we fine-tune the model on a subset of 31,000 songs in the Los Angeles dataset with a drum track. The model is asked to generate the drum track given the full score of non-percussive instruments, or to generate other instruments given a drum track.
\item \textbf{Few-shot symbolic music analysis}: we fine-tune the model on 93 songs in the RWC Pop dataset~\cite{goto2002rwc}. The model is asked to transcribe the chords and metrical structure given a symbolic pop music. We evaluate the results on symbolic chord recognition.%

\end{itemize}

In each task, we perform a random 8:1:1 split for training, validation, and testing. For the drum-to-others and others-to-drum tasks, RWC Pop is used as an external test set.

\subsection{Compared Models} \label{sec:baselines}

We compare the performance of the following models, with slight hyperparameter adjustments to ensure comparable numbers of trainable parameters.

\begin{itemize}\setlength\itemsep{0em}
    \item \textbf{FA-Cross}: The base model is fine-tuned with a cross-attentive adapter (4 heads, hidden size 256), inserted every two layers of the global decoder. A LoRA with $r=16, \alpha=32$ is used on the query and value projectors of both LMs.
    \item \textbf{FA-Self}: The base model fine-tuned with a self-attentive adapter. A LoRA with $r=64, \alpha=128$ is used on the query and value projectors of both LMs.
    \item \textbf{Coco-Mulla}: The Coco-Mulla~\cite{cocomulla} adapter applied on the RoFormer model. The adapter has a trainable positional encoding size of 384.
    \item \textbf{Prober}: A 2-layer Multilayer Perceptron (MLP) prober as used in~\cite{mert}. The MLP layer uses a weighted sum of all layers' hidden states and has a hidden dimension of 768.
    \item \textbf{Enc-Dec}: A baseline trained from scratch with a small RoFormer encoder-decoder (3 layers, hidden size 256, intermediate size 512, 4 heads for both encoder and decoder).
    \item \textbf{MelodyT5}~\cite{wu2024melodyt5}: an external baseline for the \textit{melody to chord} and \textit{chord to melody} tasks. The model is trained on 261K songs represented by ABC notations. We do not retrain this baseline.
    \item \textbf{Assistant} (Composers Assistant V2)~\cite{ComposersAssistant2}: an external baseline for the \textit{others to drum} task. We do not retrain the baseline.
\end{itemize}

All modules are trained for up to 60,000 iterations with a fixed learning rate of $10^{-4}$ and a batch size of 8 on a single A100 GPU. Early stopping is applied if validation loss \blue{does not} improve for 10 rounds (5,000 iterations).

\begin{figure}[t]
  \centering
  \includegraphics[width=\linewidth, clip, trim=0cm 0.2cm 0cm 0.2cm]{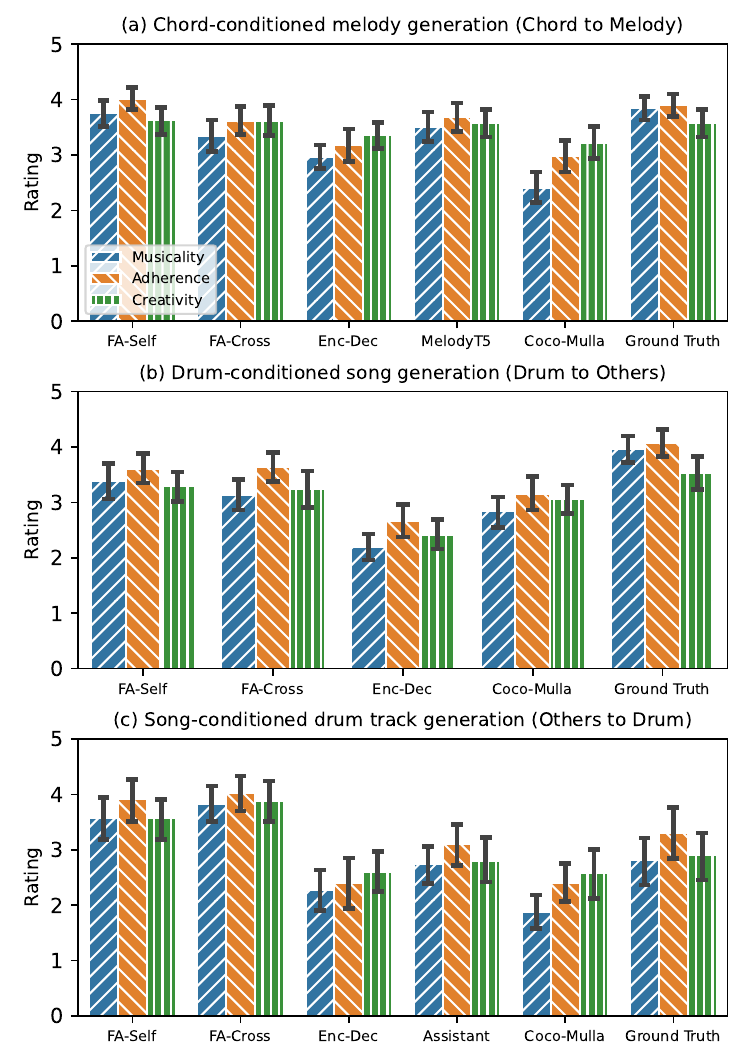}
  \caption{Subjective evaluation results. The error bars show the 95\% confidence intervals of the true mean.}
  \label{fig:subjective}
\end{figure}

\begin{table}[t] \small
\begin{tabular}{lllll}
\hline
 & \textbf{\begin{tabular}[c]{@{}l@{}}Chord to \\ melody\end{tabular}}   & \textbf{\begin{tabular}[c]{@{}l@{}}Melody to \\ chord\end{tabular}}   & \textbf{\begin{tabular}[c]{@{}l@{}}Drum to \\ others\end{tabular}}    & \textbf{\begin{tabular}[c]{@{}l@{}}Others to \\ drum\end{tabular}}    \\ \hline
FA-Cross                                               & \begin{tabular}[c]{@{}l@{}}1.4204\\ $\pm$0.0992\end{tabular}          & \begin{tabular}[c]{@{}l@{}}1.4177\\ $\pm$0.1048\end{tabular}          & \begin{tabular}[c]{@{}l@{}}2.0459\\ $\pm$0.5629\end{tabular}          & \begin{tabular}[c]{@{}l@{}}1.8619\\ $\pm$0.5665\end{tabular}          \\ \hline
FA-Self                                                & \textbf{\begin{tabular}[c]{@{}l@{}}1.4116\\ $\pm$0.1172\end{tabular}} & \textbf{\begin{tabular}[c]{@{}l@{}}1.4104\\ $\pm$0.1000\end{tabular}} & \textbf{\begin{tabular}[c]{@{}l@{}}2.0222\\ $\pm$0.6358\end{tabular}} & \textbf{\begin{tabular}[c]{@{}l@{}}1.8402\\ $\pm$0.5709\end{tabular}} \\ \hline
\begin{tabular}[c]{@{}l@{}}Coco-\\ Mulla\end{tabular}  & \begin{tabular}[c]{@{}l@{}}1.8016\\ $\pm$0.1711\end{tabular}          & \begin{tabular}[c]{@{}l@{}}1.5996\\ $\pm$0.1445\end{tabular}          & \begin{tabular}[c]{@{}l@{}}2.2027\\ $\pm$0.6532\end{tabular}          & \begin{tabular}[c]{@{}l@{}}1.9860\\ $\pm$0.6857\end{tabular}          \\ \hline
\begin{tabular}[c]{@{}l@{}}Enc-\\ Dec\end{tabular}     & \begin{tabular}[c]{@{}l@{}}1.6113\\ $\pm$0.1790\end{tabular}          & \begin{tabular}[c]{@{}l@{}}1.5067\\ $\pm$0.1208\end{tabular}          & \begin{tabular}[c]{@{}l@{}}2.5830\\ $\pm$0.9146\end{tabular}          & \begin{tabular}[c]{@{}l@{}}1.8765\\ $\pm$0.5382\end{tabular}          \\ \hline
\begin{tabular}[c]{@{}l@{}}Ground\\ Truth\end{tabular} & \begin{tabular}[c]{@{}l@{}}1.3917\\ $\pm$0.0988\end{tabular}          & \begin{tabular}[c]{@{}l@{}}1.3917\\ $\pm$0.0988\end{tabular}          & \begin{tabular}[c]{@{}l@{}}2.0730\\ $\pm$0.7158\end{tabular}          & \begin{tabular}[c]{@{}l@{}}2.0730\\ $\pm$0.7158\end{tabular}          \\ \hline
\end{tabular}
\caption{Test set perplexity on different downstream tasks.}\label{tab:ppl}
\end{table}

\subsection{Subjective Evaluation} \label{sec:subjective}

For the three generative tasks (chord-to-melody, drum-to-others, and others-to-drums), we conducted a subjective evaluation via a user survey. We selected 8 songs from the test set (2 for chord-to-melody, 4 for drum-to-others, and 2 for others-to-drums). We asked participants to rate both the generated outputs and ground truth on a 5-point scale across the following metrics:
\begin{itemize}\setlength\itemsep{0em}
\item \textbf{Musicality}: Does it sound good as music?
\item \textbf{Adherence}: Does it respect and follow the input condition?
\item \textbf{Creativity}: Given the input conditions, is it creative in its musical decisions?
\end{itemize}

We received a total of 65 answers, and the results are shown in Fig.~\ref{fig:subjective}. 

\subsection{Objective Evaluation} \label{sec:objective}

For the generative tasks by fine-tuned models, we report the generated results' perplexity on the RoFormer base model on the test set. Since perplexity is inaccurate on long repetitive generations~\cite{pplisbad}, we only calculate the perplexity using 8-bar generative results (128 steps) conditioned on 2-bar prompts (32 steps). The results are shown in Tab.~\ref{tab:ppl}.

For the melody-to-chord task, we report two additional metrics to compare with MelodyT5. We first calculate the $\textrm{L}_1$ distance between the chromagram (chroma) of the predicted chords and the ground-truth chords. We also report the CTnCTR~\cite{yeh2021automatic} metric between the melody and the generated chords. Since the test part of the Nottingham dataset has significant overlap with MelodyT5's training set, we perform a small pitch shift (up to 2 semitones) for all test songs to another commonly used key in the Nottingham dataset (e.g., C major to D major, A major to G major, etc.). The results are shown in Tab.~\ref{tab:melody-to-chord}.

For the music analysis task, we represent both the chord and the metrical labels by MIDI notes. The chord notes are represented by block notes using String Ensemble 1 (MIDI program 48). The bass note is placed in the range C3 to B3 (MIDI pitch 36-41), and other chord notes are stacked above them. We use a drum track to represent metrical labels. We use a bass drum note (MIDI pitch 35) to represent a downbeat and a snare drum note (MIDI pitch 38) for subsidiary strong beats. An 8-note infilling by closed hi-hat note (MIDI pitch 42) is also added. 

For sequence-to-sequence modeling, the model predicts both tracks from the full MIDI input, and final chord labels are derived via template matching on the average of 16 generations. The exception is the prober, trained as a 25-class classifier (12 major, 12 minor, 1 no-chord). We evaluate using chord metrics (root, majmin, seventh) from the mir\_eval package~\cite{raffel2014mir_eval}. Results are shown in Table~\ref{tab:chord}.

\subsection{Evaluation Results}\label{sec:results}

In this subsection, we analyze the results for each downstream task.

\begin{table}[t]\small
\centering
\begin{tabular}{lll}
\hline
             & \textbf{Chroma $\downarrow$}                     & \textbf{CTnCTR $\uparrow$}                     \\ \hline
Ground Truth & 0.0000$\pm$0.0000          & 0.9675$\pm$0.0324          \\ \hline
FA-Cross     & 1.5690$\pm$0.7087          & 0.9113$\pm$0.0750          \\ \hline
FA-Self      & \textbf{1.2685$\pm$0.5024} & \textbf{0.9484$\pm$0.0415} \\ \hline
Coco-Mulla~\cite{cocomulla}    & 3.4613$\pm$0.5854          & 0.6647$\pm$0.1219          \\ \hline
\blue{Enc-Dec}      & 3.0044$\pm$0.5613          & 0.8387$\pm$0.0749          \\ \hline
MelodyT5~\cite{melodyt5}     & 3.0428$\pm$0.8694          & 0.8463$\pm$0.1036          \\ \hline
\end{tabular}
\caption{Objective evaluation results on unprompted melody to chord generation on the test split of the Nottingham dataset.}\label{tab:melody-to-chord}
\end{table}

\begin{table}[t]\small
\centering
\begin{tabular}{llll}
\hline
\textbf{Model}  & \textbf{Root $\uparrow$} & \textbf{Majmin $\uparrow$} & \textbf{Seventh $\uparrow$} \\ \hline
Chorder~\cite{hsiao2021compound} & 0.7244        & 0.6760          & 0.3374           \\ \hline
\blue{HMM~\cite{wang2020pop909,jiang_midi_2025}}     & \textbf{0.8386}        & 0.8169          & 0.6930           \\ \hline
FA-Cross        & 0.8203        & 0.8455          & 0.6761           \\ \hline
FA-Self       & 0.8275        & \textbf{0.8693}          & \textbf{0.6986}           \\ \hline
Prober          & 0.8231        & 0.8370          & 0.6191           \\ \hline
\blue{Enc-Dec}         & 0.1786        & 0.1500          & 0.0378           \\ \hline
\end{tabular}
\caption{Evaluation results on symbolic chord recognition. The table shows the median result among the test split of the RWC Pop dataset.}\label{tab:chord}
\end{table}

\subsubsection{Few-shot Symbolic Music Analysis} \label{sec:symbolic_music_analysis}

With only 74 training songs, our adapters outperform rule-based baselines on both majmin and seventh categories. By comparing function alignment models (FA) with the prober, we see that using a pretrained LM for the target sequence $\mathbf{y}$ (chord+drums) improves performance on the music understanding task.

Between the function alignment models, the self-attentive adapters achieve better performance compared to cross-attentive implementation. Such trend is also observed in other tasks.

\subsubsection{Chord to Melody}

The results in subjective evaluation (Fig.~\ref{fig:subjective}a) shows that the our proposed adapters (FA-Self, FA-Cross) achieve comparable performance compared to Melody T5. Coco-Mulla is not effective on the task, achieving even lower performance compared to the Enc-Dec model. is also demonstrated in objective evaluation results (Tab.~\ref{tab:ppl}).

\subsubsection{Melody to Chord}

Both the perplexity results (Tab.~\ref{tab:ppl}) and the chord consistency results (Tab.~\ref{tab:melody-to-chord}) demonstrate the effectiveness of our models, especially the self-attentive adapters. We note that MelodyT5 shows low chroma consistency. MelodyT5 often fails to generate music that meets the constraints of the condition melody (e.g., replaced by an improvised melody or inconsistent structures). This results in a misalignment between the generated chords and the ground truth.

\subsubsection{Drum to Others}

Compared to other tasks, the drum-to-others task aims to model a highly complicated $\mathbf{y}$ (output) sequence, since $\mathbf{y}$ contains the information of the full-band arrangement. In this category, Coco-Mulla outperforms the Enc-Dec model, showing the usefulness of the pretrained knowledge from $\textrm{LM}(\mathbf{y})$. However, Coco-Mulla does not utilize the knowledge from $\textrm{LM}(\mathbf{x})$, leading to a worse performance compared to the proposed adapters.

\subsubsection{Others to Drum}

\begin{figure}
  \centering
  \includegraphics[width=\linewidth, page=10, clip, trim=0cm 6.3cm 0.2cm 0cm]{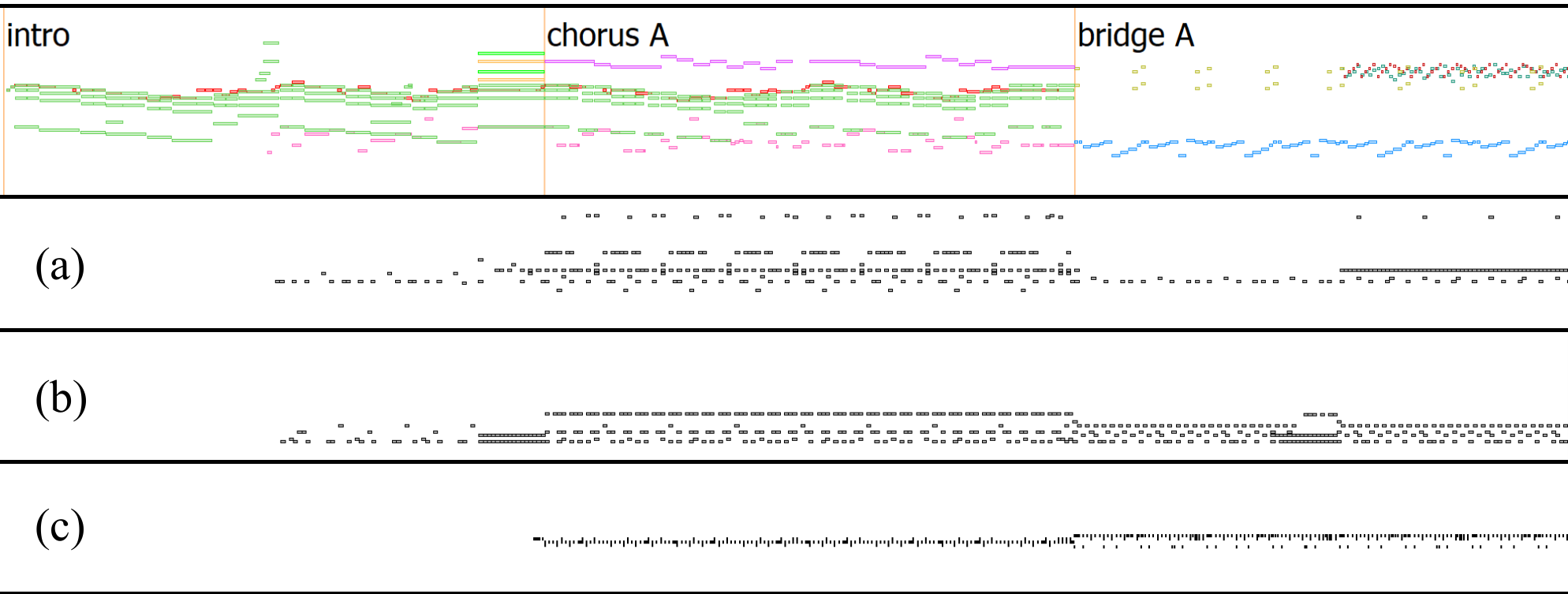}
  \caption{Case study of an others-to-drum example on RWC-Pop-003. The top displays the non-drum condition inputs with a piano roll (structure labels are shown for reference but not used by the model). The bottom shows the drum track by (a) FA-Cross; (b) FA-Self; (c) Ground truth.}
  \label{fig:case}
\end{figure}

The others-to-drum task yields the interesting results: our models outperform even the ground truth both subjectively (Fig.~\ref{fig:subjective}(c)) and objectively (Tab.~\ref{tab:ppl}). This is likely because RWC-Pop uses a limited drum set and regular patterns, while our training data (Los Angeles MIDI) includes diverse textures and instruments (e.g., Cuica). Our models generate rich, varied drum patterns aligned with long-term structure, showing strong creativity and musicality (see Fig.~\ref{fig:case} for an example). The baseline model Composer Assistant V2~\cite{ComposersAssistant2} also produces less variation.

\section{Conclusion and Future Works}

In this paper, we address the problem of versatile music-for-music modeling that unifies a broad range of music understanding and controllable generation tasks. Inspired by function alignment, we adopt a parameter-efficient approach by knowledge transfer from the pretrained LM of both the input and the output sequence. We introduce two implementations, the cross-attentive adapter and the self-attentive adapter. Both adapters show competitive results on analysis and generation tasks, with self-attentive adapters relatively outperforming.

There are mainly two future works. First, we plan to refine the data representation to support more music-for-music tasks. We also plan to extend the framework to cross-modal adapters, such as text-to-music tasks.

\bibliography{ISMIRtemplate}

%
%
%
%

\end{document}